\documentclass[11pt]{article}

\usepackage[margin=1in]{geometry}
\usepackage[T1]{fontenc}
\usepackage{lmodern}
\usepackage{microtype}
\usepackage{amsmath,amssymb}
\usepackage{booktabs}
\usepackage{tabularx}
\usepackage{enumitem}
\usepackage{xcolor}
\usepackage[hidelinks]{hyperref}
\usepackage{url}

\newcommand{\paperauthor}{Henry Zhang\\\paperaffiliation\\\texttt{github.com/henryzhangpku}}
\newcommand{\aboutauthor}{%
\paragraph{About the author.} Henry Zhang is a quantitative engineer with
fifteen years building research platforms and production trading systems,
including eight years at BlackRock and two at PIMCO. The research engine used here
is open source at
\url{https://github.com/henryzhangpku/autonomous-quant-researcher}; more
research and notes at \url{https://henryzhang.substack.com} and
\url{https://www.officialhenryzhang.com}.}

\title{Short Windows, Unmodelled Costs, Unknown Cutoffs:\\
A Pre-Registered Evidence Audit of 91 LLM Trading Papers}
\author{\paperauthor}
\date{November 2026}

\begin{document}
\maketitle

\begin{abstract}
Papers on large-language-model (LLM) trading agents and alpha mining report
Sharpe ratios that would be remarkable in any other part of empirical finance.
We audit the evidence behind those numbers. Before coding a single paper we
pre-registered a population of 91 papers from a curated corpus of LLM
trading research, a ten-field coding rubric, an analysis script and five
directional hypotheses, and committed all of them to version control. Every
code is backed by a verbatim quote that a script checks against the paper's
text, and a second, blind coder recoded a random 20\%. Of the 75 papers with a
trading test, 51\% deduct no trading costs and only 9\% model costs from data;
the median test window is 12 months, and a quarter of windows are six months or
shorter. Among the 51 papers whose test window is not known to lie after the
model's training cutoff, 98\% neither test after the cutoff, nor evaluate live,
nor use a chronologically consistent model. Only 9\% report a
multiplicity-adjusted result, and just five papers combine deducted costs,
a leakage control and a reported trial count. Two of the five hypotheses
pass (short windows; uncontrolled look-ahead). Three fail: more papers report a
trial count than we predicted, though most of those counts are repeated random
seeds rather than the number of strategies searched; and the predicted links
between headline Sharpe and test-window length, and between headline Sharpe
and costs, are not detectable in the 13 papers that state a headline Sharpe in
their abstract or conclusion. We release the codes, quotes and script, and
propose a six-line reporting minimum.
\end{abstract}

\section{Introduction}

Empirical finance learned, at some cost, that a backtest is a claim about a
search as much as about a market. Data snooping in asset pricing is an old
problem~\cite{lo1990data,white2000reality,hansen2005test}; most published
return predictors do not survive correction for the number of tests
tried~\cite{harvey2016cross,hou2020replicating}; backtest overfitting is easy to
produce and hard to detect after the
fact~\cite{bailey2014deflated,bailey2017probability}; and published
predictability decays once it is known~\cite{mclean2016does}.

LLM trading research has rebuilt much of this risk in a new form. A language
model may have read the test period during pre-training, so a backtest can
measure memory rather than reasoning~\cite{li2025timetravel,li2025mirage,gao2026lookahead}.
Systems that generate many candidate strategies select the best one, often
without reporting how many were tried~\cite{gencay2026honest}. And evaluations
are often short and narrow: when timing-based LLM strategies are re-run over two
decades and more than a hundred symbols, their reported advantage largely
disappears~\cite{li2026finsaber}.

What has been missing is a count. How many papers in this literature model
costs, control look-ahead, or report their search? The closest prior work, a
PRISMA-ScR scoping review of execution realism in LLM trading
systems~\cite{yao2026execution}, deliberately withholds its field-level counts
pending author validation. We provide those counts for a defined population,
under a design that makes the audit itself hard to bend. The population,
rubric, hypotheses and analysis script were committed before any paper was
coded; every code carries a quote that a script checks against the source text;
a blind second coder recodes a random fifth; and every change after coding is
logged.

Our results are mixed in the way a pre-registered study should allow. The
evidence base is thin on costs and look-ahead control: half of trading tests
deduct no costs, and among papers whose test may overlap the model's training
data, almost none control for it. But two of the strongest-sounding hypotheses
fail: trial counts are reported more often than we expected, and there are too
few headline Sharpe ratios in abstracts to detect the associations we
predicted.

\section{Design}

\paragraph{Population.} The corpus is a curated collection of 233 papers on LLM
trading, text-based signals, benchmarks and the classic premia and
replication literature, each with its full text extracted. The population is
every paper in themes 00--07 that is an LLM agent or alpha-mining paper, or
that reports a headline Sharpe ratio or annual return, and whose full text is
available: 91 papers. By kind, they are 50 agent papers, 13 benchmarks, 11
alpha-mining papers, 8 other methods, 5 measurement papers and 4 surveys. Most
are 2024--2026 preprints. The list was frozen before coding.

\paragraph{Rubric.} Each paper's main reported trading experiment (the one its
abstract describes) is coded on ten fields. \emph{test\_months} is the length of
the out-of-sample or test window, and \emph{n\_assets} is the number of
tradable assets. \emph{costs} is none, a flat stated rate, or realistic
(spread, impact or borrow modelled from data). \emph{leakage} is none,
acknowledged only, a test after the model's training cutoff, a live or forward
evaluation, or a chronologically consistent model~\cite{he2025chrono}.
\emph{trials} is none, mentioned, a count reported, or a multiplicity-adjusted
result (deflated Sharpe, reality check, false-discovery control). The rubric
also records a buy-and-hold baseline, the headline Sharpe as stated in the
abstract or conclusion, public code, and whether the main model's training
cutoff is known to precede the test window.

\paragraph{Hypotheses.} The directions were fixed in advance:
\begin{description}[leftmargin=*,itemsep=1pt]
\item[H1] Fewer than 10\% of papers report how many strategies or configurations
were tried, and fewer than 5\% report a multiplicity-adjusted result.
\item[H2] The median test window is 12 months or less.
\item[H3] Headline Sharpe is negatively rank-correlated with test-window length
(one-sided permutation test, $p<0.05$).
\item[H4] Among papers whose test window is not known to lie wholly after the
model's training cutoff, a majority do not address training leakage.
\item[H5] Papers that model costs report lower headline Sharpe ratios than those
that do not (one-sided permutation Mann--Whitney test, $p<0.05$).
\end{description}

\paragraph{Coding and checks.} Coding was done by language-model coders working
from the full text under the fixed rubric. Each non-empty code had to carry a
5--30-word verbatim quote. A script collapses whitespace and checks every quote
as a substring of the paper's extracted text, and none failed. A blind
second coder recoded 18 randomly chosen papers (seed fixed in advance). The
two coders agree on costs in 100\% of papers, on the buy-and-hold baseline in
94\%, on leakage, trials and code in 89\%, on test-window length (within 10\%)
in 94\%, and on the headline Sharpe (within 5\%) in 94\%. That LLMs did the
coding is itself a risk, and the quote check and second pass are there to
bound it; Section~\ref{sec:limits} discusses what they do not catch.

\paragraph{Changes after coding began.} Three, all logged. Before any result
was computed, two codes were made consistent with the rubric: one paper's live
phase was coded instead of its main backtest, and one five-day window was coded
as zero months. The first run's H4 code selected papers by their leakage code
rather than by the pre-registered cutoff criterion, which made the share 100\%
by construction. We fixed the script to match the pre-registration; no codes
changed, and the first run's output is kept.

\section{Results}\label{sec:results}

Sixteen of the 91 papers have no trading test, being benchmarks of reasoning,
forecasting accuracy, conceptual papers or reviews. The results below are for the
remaining 75 unless stated otherwise.

\begin{table}[t]
\centering
\caption{Evidence fields across 75 papers with a trading test.}\label{tab:fields}
\begin{tabular}{lrr}
\toprule
Field and value & Papers & Share \\
\midrule
\multicolumn{3}{l}{\emph{Trading costs}} \\
\quad none deducted & 38 & 51\% \\
\quad flat stated rate & 30 & 40\% \\
\quad modelled from data & 7 & 9\% \\
\multicolumn{3}{l}{\emph{Look-ahead and training leakage}} \\
\quad not discussed & 30 & 40\% \\
\quad acknowledged only & 24 & 32\% \\
\quad test after training cutoff & 14 & 19\% \\
\quad live or forward & 6 & 8\% \\
\quad chronologically consistent model & 1 & 1\% \\
\multicolumn{3}{l}{\emph{Search and multiplicity}} \\
\quad not discussed & 39 & 52\% \\
\quad mentioned & 6 & 8\% \\
\quad count reported & 23 & 31\% \\
\quad adjusted result & 7 & 9\% \\
\multicolumn{3}{l}{\emph{Other}} \\
\quad buy-and-hold or index baseline & 47 & 63\% \\
\quad public code & 22 & 29\% \\
\quad costs, leakage control and trial count together & 5 & 7\% \\
\bottomrule
\end{tabular}
\end{table}

\paragraph{Costs (Table~\ref{tab:fields}).} Half the trading tests deduct no
costs at all. Forty per cent apply a flat rate, typically 5--20 basis points
or a fixed commission, and only seven papers (9\%) model spread, impact or
borrow from data.

\paragraph{H2 passes: windows are short.} Sixty-nine papers state a test
window. The median is 12 months, with quartiles of 6 and 36 months, and 18
papers test on six months or less. The median test uses 69 assets, but 20 of 54
papers that state a count test on ten assets or fewer, often single stocks
traded one at a time.

\paragraph{H4 passes: look-ahead is rarely controlled.} For 24 papers the
main model's training cutoff is known to precede the test window, for 22 it is
known not to, and for 29 it cannot be determined from the paper. Of the 51
papers in the last two groups, 50 (98\%) neither test after the cutoff, nor
evaluate live, nor use a chronologically consistent model: their test may
overlap what the model read during training, and they either say nothing about
it (as in 40\% of all trading tests) or acknowledge it without control. Live
evaluation, the cleanest
control~\cite{li2025timetravel,chen2026nowcasting}, appears in six papers.

\paragraph{H1 fails, with a qualification.} Thirty papers (40\%) report a
trial count or an adjusted result, four times the 10\% we predicted, and seven
(9\%) report an adjusted result against a predicted 5\%. The qualification is
in what is counted. Reading the 23 quoted counts, 13 are repeated runs of one
configuration over random seeds (typically three to five), which measures run-to-run
variance, not search; ten count candidates or configurations searched, the
quantity a deflated Sharpe ratio needs. On the narrower reading, 17 papers
(23\%) disclose their search or adjust for it. The hypothesis still fails at
that reading, and we report it as failed. This split is exploratory; the
rubric did not separate the two cases.

\paragraph{H3 and H5 fail: too few headline Sharpe ratios.} Only 15 papers
state a Sharpe ratio for their method in the abstract or conclusion, and 13
after excluding relative improvements. Their median is 2.0. Within those 13,
the rank correlation between headline Sharpe and window length is $-0.10$
(one-sided permutation $p=0.36$). The cost comparison has 11 papers that model
costs and two that do not; the medians are 1.62 and 3.46, in the predicted
direction, with one-sided $p=0.10$. Neither test can detect the predicted
association at this sample size, and we do not read the cost medians as
evidence. That most papers keep their Sharpe ratio out of the abstract is
itself a finding about reporting. Many quote one in the introduction or a
table, often per asset or as gross excess.

\paragraph{Robustness.} One paper appears twice in the population, as a
preprint and as its published version. Removing the duplicate leaves every
verdict unchanged (H3: $\rho=-0.23$, $p=0.24$; H5: $p=0.10$).

\section{What the counts mean}

The pattern is consistent with a field that has adopted the vocabulary of
careful evaluation faster than its practice. A buy-and-hold baseline is
standard (63\%), and repeated seeds are common. The controls that address the
two failure modes particular to LLM backtests are rare: costs that would
survive contact with execution, and protection against a model that has read
the test period. Only five papers combine deducted costs, a leakage control and
a reported trial count, and one of those also releases code.

These are not reasons to dismiss individual papers. Several in the population
are designed around exactly these controls: live
benchmarks~\cite{li2025timetravel}, forward-logged
nowcasts~\cite{chen2026nowcasting}, leakage-safe discovery with trial-count
deflation~\cite{gencay2026honest}, production
records~\cite{barton2026production}, audits that separate an agent's selection
skill from passive exposure~\cite{chen2026audit}, and contamination
tests~\cite{gao2026lookahead}. The counts say these are exceptions.

\paragraph{A reporting minimum.} Six lines would let a reader judge any LLM
trading result in a minute:
\begin{enumerate}[leftmargin=*,itemsep=1pt]
\item the test window, dated, and the number of assets;
\item the main model and its training cutoff, and whether the test lies after it;
\item the cost model, with the breakeven cost at which the result vanishes;
\item the number of strategies, prompts or configurations searched before the
reported one;
\item a multiplicity-aware statistic (deflated Sharpe or a reality check) at that
count;
\item a benchmark-relative or composition-matched null~\cite{chen2026audit}.
\end{enumerate}
A research loop can produce these lines by construction rather than by
discipline. Our companion paper~\cite{zhang2026hardtofool} describes one that
records every trial, pre-registers each hypothesis and spends its holdout once,
in an open engine~\cite{zhang2026engine}.

\section{Limitations}\label{sec:limits}

The population is one curated corpus, not a systematic search, so the shares
describe that corpus. They are likely representative of widely circulated
work, since the corpus collects cited and recent papers, but that is not
guaranteed. The coders were language models. The quote check guarantees that
each code points to real text, not that the judgement is right; the second
pass bounds disagreement at about one paper in ten on the subjective fields,
on 18 papers only. Where a paper gives no cutoff, we code it unknown rather
than look up the model's published cutoff, which makes H4 conservative in one
direction and leaves 29 papers undetermined. The headline-Sharpe field was
defined strictly (abstract or conclusion), which kept the H3 and H5 samples
too small for their tests. A broader definition was not pre-registered, so we
do not report one. Finally, the author builds and sells research tooling in
this area; the codes, quotes and script are released so the counts can be
checked without trusting us.

\section{Conclusion}

Half the trading tests in this population deduct no costs, the median window
is a year, and almost no paper whose test may overlap its model's training
data controls for that overlap. Search counts are reported more often than we
expected, but mostly as repeated seeds rather than as the size of the search.
None of this shows that LLM trading strategies fail. It shows that most of
their published evidence could not tell us if they did.

\paragraph{Data and code.} The pre-registration (committed before coding), the
rubric, the 91 coded records with their quotes, the blind second pass, the
correction log and the analysis script are released at
\url{https://github.com/henryzhangpku/llm-trading-evidence-audit}.

\aboutauthor

\paragraph{Disclaimer.} This is a study of published research. It is not
investment advice.

\end{document}